
%
%
\def\unredoffs{} \def\redoffs{\voffset=-.31truein\hoffset=-.59truein}
\def\speclscape{}
%
%
%
%
\newbox\leftpage \newdimen\fullhsize \newdimen\hstitle \newdimen\hsbody
\tolerance=1000\hfuzz=2pt
\catcode`\@=11 
\def\bigans{b }
\def\answ{b }
%

\ifx\answ\bigans\message{(This will come out unreduced.}
\magnification=1200\unredoffs\baselineskip=.33truein plus 2pt minus 1pt
\hsbody=\hsize \hstitle=\hsize 
\else\message{(This will be reduced.} \let\l@r=L
\magnification=1000\baselineskip=16pt plus 2pt minus 1pt \vsize=7truein
\redoffs \hstitle=8truein\hsbody=4.75truein\fullhsize=10truein\hsize=\hsbody
\output={\ifnum\pageno=0 
  \shipout\vbox{\speclscape{\hsize\fullhsize\makeheadline}
   \hbox to \fullhsize{\hfill\pagebody\hfill}}\advancepageno
  \else
 \almostshipout{\leftline{\vbox{\pagebody\makefootline}}}\advancepageno
  \fi}
\def\almostshipout#1{\if L\l@r \count1=1 \message{[\the\count0.\the\count1]}
      \global\setbox\leftpage=#1 \global\let\l@r=R
 \else \count1=2
  \shipout\vbox{\speclscape{\hsize\fullhsize\makeheadline}
      \hbox to\fullhsize{\box\leftpage\hfil#1}}  \global\let\l@r=L\fi}
\fi
%
\newcount\yearltd\yearltd=\year\advance\yearltd by -1900

\def\Title#1#2{\nopagenumbers\abstractfont\hsize=\hstitle\rightline{#1}%
\vskip 1in\centerline{\titlefont #2}\abstractfont\vskip .5in\pageno=0}
\def\Date#1{\vfill\leftline{#1}\tenpoint\supereject\global\hsize=\hsbody%
\footline={\hss\tenrm\folio\hss}}
%

\def\draftmode{\message{ DRAFTMODE }\def\draftdate{{\rm preliminary draft:
\number\month/\number\day/\number\yearltd\ \ \hourmin}}%
\headline={\hfil\draftdate}\writelabels\baselineskip=20pt plus 2pt minus 2pt
 {\count255=\time\divide\count255 by 60 \xdef\hourmin{\number\count255}
  \multiply\count255 by-60\advance\count255 by\time
  \xdef\hourmin{\hourmin:\ifnum\count255<10 0\fi\the\count255}}}
\def\nolabels{\def\wrlabeL##1{}\def\eqlabeL##1{}\def\reflabeL##1{}}
\def\writelabels{\def\wrlabeL##1{\leavevmode\vadjust{\rlap{\smash%
{\line{{\escapechar=` \hfill\rlap{\sevenrm\hskip.03in\string##1}}}}}}}%
\def\eqlabeL##1{{\escapechar-1\rlap{\sevenrm\hskip.05in\string##1}}}%
\def\reflabeL##1{\noexpand\llap{\noexpand\sevenrm\string\string\string##1}}}
\nolabels
%
\global\newcount\secno \global\secno=0
\global\newcount\meqno \global\meqno=1
\def\newsec#1{\global\advance\secno by1\message{(\the\secno. #1)}
\global\subsecno=0\eqnres@t\noindent{\bf\the\secno. #1}
\writetoca{{\secsym} {#1}}\par\nobreak\medskip\nobreak}
\def\eqnres@t{\xdef\secsym{\the\secno.}\global\meqno=1\bigbreak\bigskip}
\def\sequentialequations{\def\eqnres@t{\bigbreak}}\xdef\secsym{}
\global\newcount\subsecno \global\subsecno=0
\def\subsec#1{\global\advance\subsecno by1\message{(\secsym\the\subsecno. #1)}
\ifnum\lastpenalty>9000\else\bigbreak\fi
\noindent{\it\secsym\the\subsecno. #1}\writetoca{\string\quad
{\secsym\the\subsecno.} {#1}}\par\nobreak\medskip\nobreak}
\def\appendix#1#2{\global\meqno=1\global\subsecno=0\xdef\secsym{\hbox{#1.}}
\bigbreak\bigskip\noindent{\bf Appendix #1. #2}\message{(#1. #2)}
\writetoca{Appendix {#1.} {#2}}\par\nobreak\medskip\nobreak}
%
%
\def\eqnn#1{\xdef #1{(\secsym\the\meqno)}\writedef{#1\leftbracket#1}%
\global\advance\meqno by1\wrlabeL#1}
\def\eqna#1{\xdef #1##1{\hbox{$(\secsym\the\meqno##1)$}}
\writedef{#1\numbersign1\leftbracket#1{\numbersign1}}%
\global\advance\meqno by1\wrlabeL{#1$\{\}$}}
\def\eqn#1#2{\xdef #1{(\secsym\the\meqno)}\writedef{#1\leftbracket#1}%
\global\advance\meqno by1$$#2\eqno#1\eqlabeL#1$$}
%
\newskip\footskip\footskip14pt plus 1pt minus 1pt 
\def\footnotefont{\ninepoint}\def\f@t#1{\footnotefont #1\@foot}
\def\f@@t{\baselineskip\footskip\bgroup\footnotefont\aftergroup\@foot\let\next}
\setbox\strutbox=\hbox{\vrule height9.5pt depth4.5pt width0pt}
\global\newcount\ftno \global\ftno=0
\def\foot{\global\advance\ftno by1\footnote{$^{\the\ftno}$}}
%
\newwrite\ftfile
\def\footend{\def\foot{\global\advance\ftno by1\chardef\wfile=\ftfile
$^{\the\ftno}$\ifnum\ftno=1\immediate\openout\ftfile=foots.tmp\fi%
\immediate\write\ftfile{\noexpand\smallskip%
\noexpand\item{f\the\ftno:\ }\pctsign}\findarg}%
\def\footatend{\vfill\eject\immediate\closeout\ftfile{\parindent=20pt
\centerline{\bf Footnotes}\nobreak\bigskip\input foots.tmp }}}
\def\footatend{}
%
%
\global\newcount\refno \global\refno=1
\newwrite\rfile
\def\ref{[\the\refno]\nref}
\def\nref#1{\xdef#1{[\the\refno]}\writedef{#1\leftbracket#1}%
\ifnum\refno=1\immediate\openout\rfile=refs.tmp\fi
\global\advance\refno by1\chardef\wfile=\rfile\immediate
\write\rfile{\noexpand\item{#1\ }\reflabeL{#1\hskip.31in}\pctsign}\findarg}
\def\findarg#1#{\begingroup\obeylines\newlinechar=`\^^M\pass@rg}
{\obeylines\gdef\pass@rg#1{\writ@line\relax #1^^M\hbox{}^^M}%
\gdef\writ@line#1^^M{\expandafter\toks0\expandafter{\striprel@x #1}%
\edef\next{\the\toks0}\ifx\next\em@rk\let\next=\endgroup\else\ifx\next\empty%
\else\immediate\write\wfile{\the\toks0}\fi\let\next=\writ@line\fi\next\relax}}
\def\striprel@x#1{} \def\em@rk{\hbox{}}
\def\lref{\begingroup\obeylines\lr@f}
\def\lr@f#1#2{\gdef#1{\ref#1{#2}}\endgroup\unskip}

\def\addref#1{\immediate\write\rfile{\noexpand\item{}#1}} 
\def\startrefs#1{\immediate\openout\rfile=refs.tmp\refno=#1}
\def\xref{\expandafter\xr@f}\def\xr@f[#1]{#1}
\def\refs#1{\count255=1[\r@fs #1{\hbox{}}]}
\def\r@fs#1{\ifx\und@fined#1\message{reflabel \string#1 is undefined.}%
\nref#1{need to supply reference \string#1.}\fi%
\vphantom{\hphantom{#1}}\edef\next{#1}\ifx\next\em@rk\def\next{}%
\else\ifx\next#1\ifodd\count255\relax\xref#1\count255=0\fi%
\else#1\count255=1\fi\let\next=\r@fs\fi\next}
%

%
\newwrite\ffile\global\newcount\figno \global\figno=1
\def\fig{Figure~\the\figno\nfig}
\def\nfig#1{\xdef#1{Figure~\the\figno}%
\writedef{#1\leftbracket fig.\noexpand~\the\figno}%
\ifnum\figno=1\immediate\openout\ffile=figs.tmp\fi\chardef\wfile=\ffile%
\immediate\write\ffile{\noexpand\medskip\noexpand\item{Fig.\ \the\figno. }
\reflabeL{#1\hskip.55in}\pctsign}\global\advance\figno by1\findarg}
\def\xfig{\expandafter\xf@g}\def\xf@g fig.\penalty\@M\ {}
\def\figs#1{figs.~\f@gs #1{\hbox{}}}
\def\f@gs#1{\edef\next{#1}\ifx\next\em@rk\def\next{}\else
\ifx\next#1\xfig #1\else#1\fi\let\next=\f@gs\fi\next}
\newwrite\lfile
{\escapechar-1\xdef\pctsign{\string\%}\xdef\leftbracket{\string\{}
\xdef\rightbracket{\string\}}\xdef\numbersign{\string\#}}

\def\writestop{\def\writestoppt{\immediate\write\lfile{\string\pageno%
\the\pageno\string\startrefs\leftbracket\the\refno\rightbracket%
\string\def\string\secsym\leftbracket\secsym\rightbracket%
\string\secno\the\secno\string\meqno\the\meqno}\immediate\closeout\lfile}}
\def\writestoppt{}\def\writedef#1{}
\def\seclab#1{\xdef #1{\the\secno}\writedef{#1\leftbracket#1}\wrlabeL{#1=#1}}
\def\subseclab#1{\xdef #1{\secsym\the\subsecno}%
\writedef{#1\leftbracket#1}\wrlabeL{#1=#1}}
\newwrite\tfile \def\writetoca#1{}
\def\leaderfill{\leaders\hbox to 1em{\hss.\hss}\hfill}
\def\writetoc{\immediate\openout\tfile=toc.tmp
   \def\writetoca##1{{\edef\next{\write\tfile{\noindent ##1
   \string\leaderfill {\noexpand\number\pageno} \par}}\next}}}

%
\catcode`\@=12 
%
\edef\tfontsize{\ifx\answ\bigans scaled\magstep3\else scaled\magstep4\fi}
\font\titlerm=cmr10 \tfontsize \font\titlerms=cmr7 \tfontsize
\font\titlermss=cmr5 \tfontsize \font\titlei=cmmi10 \tfontsize
\font\titleis=cmmi7 \tfontsize \font\titleiss=cmmi5 \tfontsize
\font\titlesy=cmsy10 \tfontsize \font\titlesys=cmsy7 \tfontsize
\font\titlesyss=cmsy5 \tfontsize \font\titleit=cmti10 \tfontsize
\skewchar\titlei='177 \skewchar\titleis='177 \skewchar\titleiss='177
\skewchar\titlesy='60 \skewchar\titlesys='60 \skewchar\titlesyss='60
\def\titlefont{\def\rm{\fam0\titlerm}
\textfont0=\titlerm \scriptfont0=\titlerms \scriptscriptfont0=\titlermss
\textfont1=\titlei \scriptfont1=\titleis \scriptscriptfont1=\titleiss
\textfont2=\titlesy \scriptfont2=\titlesys \scriptscriptfont2=\titlesyss
\textfont\itfam=\titleit \def\it{\fam\itfam\titleit}\rm}
 \ifx\answ\bigans\else scaled\magstep1\fi
\ifx\answ\bigans\def\abstractfont{\tenpoint}\else
\font\abssl=cmsl10 scaled \magstep1
\font\absrm=cmr10 scaled\magstep1 \font\absrms=cmr7 scaled\magstep1
\font\absrmss=cmr5 scaled\magstep1 \font\absi=cmmi10 scaled\magstep1
\font\absis=cmmi7 scaled\magstep1 \font\absiss=cmmi5 scaled\magstep1
\font\abssy=cmsy10 scaled\magstep1 \font\abssys=cmsy7 scaled\magstep1
\font\abssyss=cmsy5 scaled\magstep1 \font\absbf=cmbx10 scaled\magstep1
\skewchar\absi='177 \skewchar\absis='177 \skewchar\absiss='177
\skewchar\abssy='60 \skewchar\abssys='60 \skewchar\abssyss='60
\def\abstractfont{\def\rm{\fam0\absrm}
\textfont0=\absrm \scriptfont0=\absrms \scriptscriptfont0=\absrmss
\textfont1=\absi \scriptfont1=\absis \scriptscriptfont1=\absiss
\textfont2=\abssy \scriptfont2=\abssys \scriptscriptfont2=\abssyss
\textfont\itfam=\bigit \def\it{\fam\itfam\bigit}\def\footnotefont{\tenpoint}%
\textfont\slfam=\abssl \def\sl{\fam\slfam\abssl}%
\textfont\bffam=\absbf \def\bf{\fam\bffam\absbf}\rm}\fi
\def\tenpoint{\def\rm{\fam0\tenrm}
\textfont0=\tenrm \scriptfont0=\sevenrm \scriptscriptfont0=\fiverm
\textfont1=\teni  \scriptfont1=\seveni  \scriptscriptfont1=\fivei
\textfont2=\tensy \scriptfont2=\sevensy \scriptscriptfont2=\fivesy
\textfont\itfam=\tenit \def\it{\fam\itfam\tenit}\def\footnotefont{\ninepoint}%
\textfont\bffam=\tenbf \def\bf{\fam\bffam\tenbf}\def\sl{\fam\slfam\tensl}\rm}
\font\ninerm=cmr9 \font\sixrm=cmr6 \font\ninei=cmmi9 \font\sixi=cmmi6
\font\ninesy=cmsy9 \font\sixsy=cmsy6 \font\ninebf=cmbx9
\font\nineit=cmti9 \font\ninesl=cmsl9 \skewchar\ninei='177
\skewchar\sixi='177 \skewchar\ninesy='60 \skewchar\sixsy='60
\def\ninepoint{\def\rm{\fam0\ninerm}
\textfont0=\ninerm \scriptfont0=\sixrm \scriptscriptfont0=\fiverm
\textfont1=\ninei \scriptfont1=\sixi \scriptscriptfont1=\fivei
\textfont2=\ninesy \scriptfont2=\sixsy \scriptscriptfont2=\fivesy
\textfont\itfam=\ninei \def\it{\fam\itfam\nineit}\def\sl{\fam\slfam\ninesl}%
\textfont\bffam=\ninebf \def\bf{\fam\bffam\ninebf}\rm}
%
%

\hyphenation{anom-aly anom-alies coun-ter-term coun-ter-terms}
\def\inv{^{\raise.15ex\hbox{${\scriptscriptstyle -}$}\kern-.05em 1}}

\def\Dsl{\,\raise.15ex\hbox{/}\mkern-13.5mu D} 
\def\dsl{\raise.15ex\hbox{/}\kern-.57em\partial}

 \def\Tr{{\rm Tr}}
\font\bigit=cmti10 scaled \magstep1
\def\lspace{\ifx\answ\bigans{}\else\qquad\fi}
\def\lbspace{\ifx\answ\bigans{}\else\hskip-.2in\fi} 
\def\boxeqn#1{\vcenter{\vbox{\hrule\hbox{\vrule\kern3pt\vbox{\kern3pt
	\hbox{${\displaystyle #1}$}\kern3pt}\kern3pt\vrule}\hrule}}}
\def\mbox#1#2{\vcenter{\hrule \hbox{\vrule height#2in
		\kern#1in \vrule} \hrule}}  
%

\def\darr#1{\raise1.5ex\hbox{$\leftrightarrow$}\mkern-16.5mu #1}

\def\half{{\textstyle{1\over2}}} 
\def\roughly#1{\raise.3ex\hbox{$#1$\kern-.75em\lower1ex\hbox{$\sim$}}}

\def\perpp{{\scriptscriptstyle\perp}}

\def\half{{1\over 2}}

\def\free{\hbox{$\cal F$}}

\def\bo#1{{\cal O}(#1)}

\def\free{\hbox{$\cal F$}}
\def\bold#1{\setbox0=\hbox{$#1$}%
     \kern-.010em\copy0\kern-\wd0
     \kern.025em\copy0\kern-\wd0
     \kern-.020em\raise.0200em\box0 }

\def\dz{\partial_z}

\def\ts{\theta_6}

\def\cross{\!\times\!}
\def\dnb{\delta {\vec n}}
\def\dot{\!\cdot\!}

\nfig\fphase{Phase diagram of a chiral polymer crystal.  Insets are
representative
tilt (TGB) and moir\'e grain boundaries.  Shaded lines are screw dislocations.}

\nfig\frandom{A projected top view of the moir\'e map with rotation angle
$\phi_1\approx21.8^{\circ}$ iterated three times.  The inset shows $20$ random
polymers paths resulting from the same map iterated 99 times.}

\nfig\fcomplex{The moir\'e state.  The thick tubes
running in the $\hat z$ direction are polymers, while the dark lines are
stacked honeycomb arrays of screw dislocations.}

\lref\MEYER{R.B.~Meyer, {\sl Polymer Liquid
Crystals}, edited by A. Ciferri, W.R. Kringbaum and R.B. Meyer (Academic, New
York, 1982) Chapter 6.}
\lref\MN{M.C.~Marchetti and D.R.~Nelson, Phys. Rev. B {\bf 41} 1910 (1990).}
\lref\TGB{S.R.~Renn and T.C.~Lubensky, Phys. Rev. A {\bf 38}, 2132 (1988);
G.~Srajer, R.~Pindak, M.A.~Waugh and
J.W.~Goodby, Phys. Rev. Lett. {\bf 64}, 1545 (1990); see also
P.G.~de Gennes, Solid State Commun. {\bf 10}, 753 (1972).}

\lref\TON{J.~Toner, Phys. Rev. A {\bf 27} 1157 (1983).}
\lref\DGP{P.G.~de Gennes and J.~Prost, The Physics of Liquid Crystals, Second
ed.,
(Oxford University Press, New York, 1993).}

\lref\KO{M.~Kl\'eman and P.~Oswald, J. Phys. (Paris) {\bf 43} 655 (1982).}

\lref\BOUii{F.~Livolant
and Y.~Bouligand, J. Phys. (Paris) {\bf 47} 1813 (1986); For tilt and moir\'e
boundaries in the non-biological polymer PBZO (poly-paraphenylene
benzobisoxazole) see
D.C.~Martin and E.L.~Thomas, Phil. Mag. A {\bf 64}, 903 (1991).}
\lref\PARS{R.~Podgornik and V.A.~Parsegian, Macromolecules {\bf 23} 2265
(1990).}
\lref\GIA{C.~Gianessi, Phys. Rev. A {\bf 28} 350 (1983); Phys. Rev. A {\bf 34}
705 (1986).}
\lref\NT{D.R.~Nelson and J.~Toner, Phys. Rev. B {\bf 24} 363 (1981), and
references
therein.}
\lref\IND{V.L.~Indenbom and A.N.~Orlov, Usp. Fiz. Nauk {\bf 76} 557 (1962) [
Sov. Phys. Uspekhi {\bf 5} 272 (1962)].}
\lref\HN{D.R.~Nelson and J.~Toner, Phys. Rev. B {\bf 24}, 363 (1981) and
references therein.}
\lref\TER{E.M.~Terentjev, Europhys. Lett. {\bf 23}, 27 (1993).}
\lref\KN{R.D.~Kamien and D.R.~Nelson, unpublished.}

\Title{IASSNS-HEP-94/68}{\vbox{\centerline{Iterated Moir\'e Maps and}\vskip2pt
\centerline{Braiding of Chiral Polymer Crystals}}}

\centerline{Randall D. Kamien\foot{email: kamien@guinness.ias.edu}}
\centerline{\sl School of Natural Sciences, Institute for Advanced Study,
Princeton, NJ
08540}
\centerline{and}
\centerline{David R. Nelson}
\centerline{\sl Lyman Laboratory of Physics,
Harvard University, Cambridge, MA 02138}

\vskip .3truein
In the hexagonal columnar phase of chiral polymers a bias towards cholesteric
twist competes with braiding along an average direction.
When the chirality is strong,
screw dislocations proliferate, leading to either a tilt grain boundary
phase or a new
``moir\'e state'' with twisted bond order. Polymer
trajectories in the plane perpendicular to their average direction are
described by
iterated moir\'e maps of remarkable complexity.

\Date{9 November 1994}

A notable feature of biological materials is the profusion of long polymer
molecules
with a definite handedness.  DNA, polypeptides (such as
poly-$\gamma$-benzyl-glutamate)
and polysaccharides (such as xanthan) can all be
synthesized with a preferred chirality.
Long polymers in dense solution often crystallize into a hexagonal columnar
phase.
When the polymers
are chiral this close packing into a triangular lattice
competes with the tendency for the polymers to twist macroscopically \BOUii\
as in cholesteric liquid crystals.  Similar to the twist grain boundary
phase of chiral smectics \TGB , macroscopic chirality can proliferate when
screw
dislocations enter the crystal.
Like flux lines in
a type II superconductor, dislocations only appear provided the free energy
reduction
from the chiral couplings
exceeds the dislocation core energy.  If the chirality is weak, a defect
free hexagonal columnar phase persists, as in the Meissner phase of
superconductors.

In this letter we explore the effect of chirality
on the hexagonal columnar phases \DGP\ of long
polymers in detail.
We neglect for simplicity heterogeneity along the polymer backbones and work
with a two component displacement field perpendicular to the local polymer
direction.
The
usual chiral term relevant for cholesteric liquid crystals produces the polymer
tilt grain
boundary phase,
similar to the
smectic-$A^*$ phase \TGB .  We find, as well, an additional
term in the free energy which favors the rotation of the bond order along the
average
polymer direction.
This term leads to braided polymers with twisting describable
by a sequence of moir\'e patterns.

Using continuum elastic theory we estimate the critical values
of the chiral couplings above which screw dislocations enter.  On a more
microscopic level, we propose a set of lock-in moir\'e textures
which should be especially low in elastic energy.  These states are
entangled and self-similar, and lead to tortuous polymer paths of
remarkable complexity, reminiscent of chaotic dynamical systems.

When polymer nematics crystallize
the areal polymer density in a plane perpendicular to the average direction
may be approximated as
\eqn\waves{\rho \approx \rho_0 + \sum_\alpha \rho_\alpha(
{\vec r}) \exp\{-i{\vec G_\alpha}\dot{\vec r}\}}
where the $\{{\vec G}_\alpha\}$ are the six smallest
reciprocal lattice vectors of a triangular lattice.
We take the average polymer
direction here and throughout to be $\hat z$, with $\vec G_\alpha\dot\hat z
=0$.
Each plane wave is modulated by a spatially varying magnitude and
phase $\rho_\alpha({\vec r}) = \vert\rho_\alpha({\vec r})\vert{\exp\{i{\vec
G_\alpha}\dot
{\vec u}({\vec r})\}}$, where $\vec u$ is a two-dimensional displacement field.
In addition, we have a nematic order parameter, $\hat n\approx
\hat z + \dnb$ and, in the case of a triangular lattice of polymers, a bond
field $\ts$
which measures
the bond-orientational order in the $xy$-plane, modulo $2\pi/6$.
Under a global rotation about the $x$-axis or
$y$-axis by an angle $\theta_x$ or $\theta_y$ respectively, $\hat n\rightarrow
\hat n + \vec\theta\cross\hat n\approx \hat z +\theta_y\hat x -\theta_x\hat y$,
{\sl
i.e.} $\dnb \rightarrow \dnb + \theta_y\hat x -\theta_x\hat y$.  Similarly
under a global rotation about the $z$-axis by $\theta_z$,
$\ts\rightarrow\ts+\theta_z$.
Under such rotations,
$\vec r\rightarrow\vec r-\vec\theta\cross\vec r$,
leading to a position dependent change in the phase of $\rho_\alpha(\vec r)$.
To insure rotational invariance, derivatives of $\rho_\alpha(\vec r)$
must be accompanied by the fields $\ts$ and $\dnb$ \HN\ and we are led to the
Landau
free energy
\eqn\efree{\eqalign{
\free_{\rm d}&=\sum_\alpha\bigg\{
{A\over 2}\left\vert\vec G_\alpha\dot\left[
\nabla_{\!\perpp}\rho_\alpha - i\ts\left(\vec
G_\alpha\cross\hat z\right)\rho_\alpha\right]\right
\vert^2
+{B\over 2}\left\vert\vec G_\alpha\cross\left[
\nabla_{\!\perpp}\rho_\alpha - i\ts\left(\vec
G_\alpha\cross\hat z\right)\rho_\alpha\right]\right
\vert^2
\cr
&\quad+{C\over 2}\left\vert
\partial_z\rho_\alpha - i\left(\vec
G_\alpha\dot\dnb\right)\rho_\alpha\right\vert^2
+ {b\over 2}\vert\rho_\alpha\vert^2\bigg\}+c\sum_{\alpha\beta\gamma}^{\vec
G_\alpha
+\vec G_\beta +\vec G_\gamma = \vec 0}
\rho_{\alpha}\rho_{\beta}\rho_{\gamma} +
\bo{\rho_\alpha^4}\cr}}
Crystalline order arises for $b$ sufficiently negative so
that
$\langle\,\rho_\alpha\,\rangle\ne 0$ .  Due
to the third order term, this transition will, in general, be first order.

To this free energy we add two more pieces.
It is convenient, but not essential, to imagine that the hexagonal columnar
density waves arise from a phase with local nematic order as well as six-fold
bond order perpendicular to the director axis.  This ``N+6'' phase has been
studied by
Toner \TON .  Although the present experimental evidence for N+6 order in
polymer
nematics is sketchy, such phases seem highly likely in columnar systems, in
analogy with
the hexatic order expected for vortex lines in high temperature superconductors
\MN .
In the absence of chirality we have the usual elastic energies of the nematic
field and
the bond angle field,
\eqn\efreeiv{\free_{\rm o} = \half \left[K_1(\nabla_{\!\perpp}\dot\dnb)^2
+K_2(\nabla_{\!\perpp}\cross\dnb)^2 +
K_3(\partial_z\dnb)^2 + K_A^{||}(\partial_z\ts)^2 + K_A^\perp
(\nabla_{\!\perpp}\ts)^2\right],}
where the $\{K_i\}$ are Frank constants, and $K_A^{||}$ and $K_A^\perp$ are
hexatic stiffnesses parallel and perpendicular to $\hat z$.  For long chain
polymers
$K_1\gg K_2, K_3$ \MEYER.

Finally, we add those chiral terms which respect the nematic symmetry.  If
there is no
preferred direction along the polymers then the free energy must be invariant
under $\hat
n\rightarrow -\hat n$.  Additionally, since changes in $\ts$ are measured with
respect
to the $\hat n$ axis, under nematic inversion, $\ts\rightarrow -\ts$, and the
vector
$\vec v\equiv\nabla\ts$ changes sign.  Under {\sl spatial} inversion
$\hat n\rightarrow -\hat n$, $\ts\rightarrow -\ts$ and hence $\vec v\rightarrow
\vec v$.
The chiral free energy density $\free^*$ contains two distinct terms
invariant under nematic inversion but which
change sign under parity \TER ,
\eqn\echir{\free^* = -K_2q_0\hat n\dot\left(\nabla\cross\hat n\right) -
K_A^{||}\tilde q_0\vec v\dot\hat n \approx -\gamma\nabla_{\!\perpp}\cross\dnb
-\gamma'\partial_z\ts}
with $\gamma = K_2q_0$ and $\gamma'=K_A^{||}\tilde q_0$.
The first term often leads to cholesteric twist with
period $2\pi/q_0$ along a line perpendicular to the plane in which the
nematic director lies.  The second induces twist into the
bond-order parameter $\ts$ with period $2\pi/\tilde q_0$ along $\hat z$.  The
consequences of this additional chiral coupling is the main subject of this
paper.
The total free energy is $F=\int d^3\!r\,\left[\free_{\rm d} + \free_{\rm o} +
\free^*\right]$.  Note the close similarity between \efree-\echir\
and the Ginsburg-Landau
theory of a superconductor in a magnetic field.  The fields $\dnb$ and $\ts$
are ``gauge
fields'' minimally
coupled to the complex order parameters $\{\rho_\alpha(\vec r)\}$ by the
constraint
of rotational invariance.
There are {\sl two} distinct ``magnetic'' fields $\gamma$ and $\gamma'$.

Provided $b\ll 0$ we can set $\rho_\alpha = \vert\rho_0\vert\exp\{i\vec
G_\alpha\dot\vec
u(\vec r)\}$ and minimize $F$ to find
$\delta n_i = \partial_z u_i$ and
$\ts=\half\epsilon_{ij}\partial_iu_j$.  The resulting elastic free energy is
now
\eqn\efreelast{F=\int d^3\!r\,\left\{\mu u_{ij}^2 + {\lambda\over 2}u_{ii}^2 +
K_3(\partial_z^2u_i)^2 -\gamma\nabla_{\!\perpp}\cross\dnb
-\gamma'\partial_z\ts\right\} }
where $u_{ij} =\half(\partial_iu_j+\partial_ju_i)$,
$\mu = {3\over 4}\vert\vec G\vert^4\vert\rho_0\vert^2(A+B)$ and
$\lambda={3\over 4}\vert\vec G\vert^4\vert\rho_0\vert^2(A-B)$. In the columnar
crystal the
two chiral terms are the same if $\partial_z\partial_i=\partial_i\partial_z$.
However, in the presence of dislocations derivatives do {\sl not} commute,
$\nabla_{\!\perpp}\cross\dnb = \epsilon_{ij}\partial_i\partial_z
u_j\ne\partial_z
\epsilon_{ij}\partial_iu_j = \partial_z\ts$.  Here, and throughout, $i,j\ldots$
refer to
indices in the $xy$-plane.
Burgers vectors of dislocations in a hexagonal columnar phase
lie in the $xy$-plane, and there are three generic types \DGP :
a screw dislocation, an edge dislocation
with tangent along $\hat z$, and an edge dislocation lying in the
$xy$-plane.  The latter defect requires aligned polymer ends which we neglect
in
this paper.  The remaining dislocations must lie in a plane spanned by their
Burgers
vector $\vec b$ and $\hat z$, which amounts to choosing dislocation complexions
with
$\alpha_{xy}=\alpha_{yx}$, where the dislocation density tensor $\alpha_{\gamma
i}$ is the
density of dislocations with tangents
along the $\gamma$-direction with Burgers vectors pointing
in the $i$ direction \MN.

Proceeding as
in \refs{\IND,\MN}\ we introduce a new field $w_{\gamma i} \equiv
\partial_\gamma u_i$
away from any dislocations.  The non-commutivity of
the derivatives of $\vec u$ is represented by the dislocation density,
$\epsilon_{\mu\nu\gamma}
\partial_\nu
w_{\gamma i} = -\alpha_{\mu i}$.
One can solve for $w_{\gamma i}$ in terms of the dislocation density
$\alpha_{\gamma i}$ and find the equilibrium displacement field in the presence
of
crystal defects \KN . In terms
of the nematic and bond order field non-commutivity of derivatives means
$2\dz\ts-\nabla_{\!\perpp}\cross\dnb =
-\Tr[\alpha]$.  The energy per unit length
of a screw dislocation is finite,
while that of an edge dislocation lying along the $\hat z$ direction diverges
logarithmically with system size \KO .
The screw dislocation energy
depends on the dimensionless parameter
$\delta=(K_3\xi_\perpp^2/\mu\xi_z^4)^{1/4}$, where
$\xi_\perpp$ and $\xi_z$ are short distance cutoffs in the perpendicular
and parallel directions, respectively.  We estimate
$\delta\gg 1$ for polymer crystals which is the condition that the response
to chirality be similar to that of Type II as opposed to Type I superconductors
\TGB .
In this limit the free energy per unit length of a screw dislocation is
$f_{\rm s} = (\mu^{3/4}K_3^{1/4}b^2)/(2\pi\sqrt{2\xi_\perpp})$,
where $b$ is the length of the Burgers vector.

If the chirality is strong and $\gamma\gg\gamma'$ we expect
the polymer analogue of the Renn-Lubensky
twist-grain-boundary state \TGB . Each of these tilt grain boundaries (TGB)
is composed of a parallel array of
screw dislocations lying, say,
in the $xz$-plane, pointing along the $x$-axis and uniformly spaced along $\hat
z$ with spacing
$d$.
As illustrated in \fphase\ this dislocation texture causes a discrete rotation
$\phi=\tan^{-1}(b/d)$ in
the average polymer direction.
The spatial integral of $\nabla_{\!\perpp}\cross\dnb$ is non-zero, while the
integral of $\dz\ts$ vanishes.  The TGB state appears when the
chiral coupling $\gamma$ exceeds the critical value $\gamma_c=f_{\rm screw}/b$.
The diffraction signature is similar to that of the smectic-$A^*$ phase:
if the
pitch axis is $\hat y$ so that the polymers lie, on average, in the $xz$-plane,
the structure function would consist of two Bragg circles with radius $2\pi/a$
lying
in a $q_xq_z$-plane and centered on $\vec q=(0,\pm2\pi/\sqrt{3}a,0)$
where $a_0$ is the lattice constant and two Bragg spots at $q_y=\pm
4\pi/\sqrt{3}a_0$.
These features should have widths $\sim 2\pi/d'$ along $\hat q_y$ where $d'$ is
the
spacing between tilt grain boundaries.
If the rotation angle between successive crystalline regions were a rational
fraction of
$2\pi$, the Bragg circles would break up into a sequence of Bragg spots spaced
around the
circle.  Although a precise determination of $d$ and $d'$ would
require more detailed energetic calculations, we estimate, following \TGB ,
that
$d\approx d'\approx\sqrt{a_0/q_0}$.

The second chiral coupling $\gamma'$ has no analogue in chiral smectics.  To
find a
configuration of screw dislocations which exploits this form of chirality
we search for a dislocation texture which produces a displacement
depending only on $z$. The only texture which does
not produce divergent elastic energy is one in which the only
non-zero
components of $\alpha_{\gamma i}$ are $\alpha_{xx}=\alpha_{yy}$ \refs{\MN,\KN}.
A honeycomb array of
screw dislocations, on average, produces this dislocation texture.
We
now find
that for a single such grain boundary $\nabla_{\!\perpp}\cross\dnb$ vanishes
far from
the boundary while
$\partial_z\ts$ does not. This sort of grain boundary thus causes a net twist
of
the hexatic order parameter $\ts$ while imposing no net
$\nabla_{\!\perpp}\cross\dnb$.
When $\gamma'\gg\gamma$ and the chirality is large, screw dislocations
penetrate
for $\gamma'> \gamma'_c = 2f_{\rm s}/b$.  By choosing the rotation angle across
the
honeycomb dislocation network to produce a high density of coincidence lattice
sites \ref\BOL{
W.A.~Bollman, Crystal Defects and Crystalline Interfaces, (Springer-Verlag,
Berlin, 1970);
J.P.~Hirth and J.~Lothe, Theory of Dislocations, Second ed. (Wiley, New York,
1982).},
we produce
especially low strain energies across the boundary. The superposition of
triangular polymer
lattices below and above the boundary forms a moir\'e pattern.  The
superposition of many such ``moir\'e sheets'' along the $z$-axis braids
the polymers
with deep minima in the energy at
certain lock-in angles.
Figure 1 illustrates the mapping of polymers across the moir\'e plane, for
a rotation angle $\phi_2\approx 13.2^{\circ}$ with a particularly dense set of
coincidence lattice sites.  Polymers
in
the lower half-space (circles) must be connected to the closest available
polymer in the
upper half-space (crosses) to minimize bending energy.
Note that the map has a discrete translational
symmetry, in the sense that any coincidence site could be a center of rotation.
 Especially
simple moir\'e maps arise for rotation angles
$\phi_n = 2\,\tan^{-1}[{\sqrt{3}/ 3(2n+1)}]$,
$n=1,2,\ldots$. It can be shown that all such angles are irrational fractions
of
$2\pi$ \KN\ so that the structure never repeats upon iteration.
Around each coincidence point
there are $n$ concentric rings of helical polymers  The lattice of coincidence
points is
also a triangular lattice, but with a spacing $a_n=a_0\sqrt{1+3(2n+1)^2}/2$,
where $a_0$ is the original lattice spacing.
The geometrical origin of such energetically preferred lock-in angles has no
analogue
in chiral smectics.  The exact choice of lock-in angles and spacing between
moir\'e planes
must again be settled by detailed energetic calculations.

Upon two iterations of the moir\'e map separating three regions of
polymer crystal, the first coincidence lattice is rotated with respect to
the second coincidence lattice by precisely the angle of rotation $\phi_n$.
Thus the composite coincidence
lattice is the ``coincidence lattice of coincidence lattices'', with lattice
constant $a_n^2/a_0$.   Moir\'e maps iterated $p$ times lead to triangular
composite
coincidence lattices with spacing $a_n(a_n/a_0)^{p-1}$, {\sl i.e.} to
ever sparser lattices of fixed points with intricate fractal
structure in between them.  Figure 2 shows the projected polymer
paths for $p=3$ and $n=1$.  In contrast to
the
TGB state polymers for $p\gg 1$ are
highly entangled and wander far from straight line trajectories.
The polymer configurations and the
dislocations leading to them are shown in \fcomplex\ for $n=1$.  Near the
moir\'e planes both $\ts$ and $\nabla_{\!\perpp}\cross\dnb$ are nonzero.
The center polymer is
a fixed point of all the maps.  Any such fixed polymer has
a halo of others twisting around it.  In this special tube, the
nematic order parameter takes on the texture of a double twist cylinder as
found
in the low-chirality limit of
blue phases \ref\SWM{D.C.~Wright and N.D.~Mermin, Rev. Mod. Phys. {\bf 61}, 385
(1989).}.
The moir\'e state takes advantage of both double twist energies and the
new chiral coupling $\gamma'$.  Closely related states are possible in the
chiral N+6 phases.  Textures of {\sl vortex lines} in Type II superconductors,
similar
to Figure 3, may arise when the Abrikosov flux lattice is subjected to a strong
supercurrent parallel to the field direction \KN .

In the inset to \frandom\ we show a random selection of polymer paths projected
onto the $xy$-plane
representing $99$ iterations of the moir\'e map.  Though these polymers are
clearly
influenced by
the exceptional center point which is a fixed point of all $99$ mappings, there
are
still paths of great complexity such that polymers wander far from their
original positions.
In a defect-free hexagonal columnar phase the most intense Bragg spots appear
in the $q_z=0$ plane
at the vertices of a hexagon with radius $G= 4\pi/\sqrt{3}a_0$.
The braided crystal described by the iterated moir\'e map should exhibit a
Bragg ring at $\vert \vec q_{\perpp}\vert=G$ in the $q_z=0$ plane.
The ring should be very sharp along $\vec q_\perp$, and of width $\sim 2\pi/d'$
along
$\hat q_z$, where $d'$ is the spacing between moir\'e planes.  We estimate
that $d'\approx a_n\approx \sqrt{a_0/\tilde q_0}$.  As in the Renn-Lubensky
state \TGB ,
rational moir\'e rotations with angle $\phi=2\pi s/t$, where $s$ and $t$ are
relatively
prime integers,
would cause the Bragg rings to
break up into $t$ spots spaced out around the ring, with additional structure
along
$\hat q_z$.

The phase diagram in \fphase\ summarizes our conclusions.  The hexagonal
columnar
phase is like the Meissner phase of Type II superconductors.  The {\sl two}
chiral couplings
$\gamma$ and $\gamma'$ cause screw dislocations to penetrate the crystal above
critical strengths $\gamma_c$ and $\gamma_c'$, in much the same way as vortices
enter
Type II superconductors above the lower critical field $H_{\rm c1}$.
The TGB phase predicted here for chiral polymers is similar to that already
observed
experimentally in chiral smectics \TGB .  The braided moir\'e state is
qualitatively
new, and its experimental observation would be of considerable interest.
Surprising
little is known about the intricate trajectories produced by iterated moir\'e
maps
shown in \frandom .  These could be studied experimentally via neutron
diffraction in hexagonal
columnar crystals with a dilute concentration of deuterated polymer strands.
Numerical studies of the local fractal dimension and Lyapunov exponent are
currently
in progress.

It is a pleasure to acknowledge stimulating conversations with T.~Lubensky,
R.~Meyer,
P.~Taylor, E.~Thomas,
and J.~Toner.
RDK was supported by National Science Foundation Grant
No.~PHY92--45317.  DRN acknowledges the hospitality of Brandeis University,
AT\&T Bell
Laboratories, and Exxon Research and Engineering, as well as support from the
Guggenheim
Foundation and the National Science Foundation, through the Harvard Materials
Research Laboratory and Grant No. DMR-91-15491  .
\footatend\vfill\supereject\immediate\closeout\rfile\writestoppt
\baselineskip=.33truein\centerline{{\bf References}}\bigskip{\frenchspacing%
\parindent=20pt\escapechar=` \input refs.tmp\vfill\eject}\nonfrenchspacing
\vfill\eject\immediate\closeout\ffile{\parindent40pt
\baselineskip.33truein\centerline{{\bf Figure Captions}}\nobreak\medskip
\escapechar=` \input figs.tmp\vfill\eject}

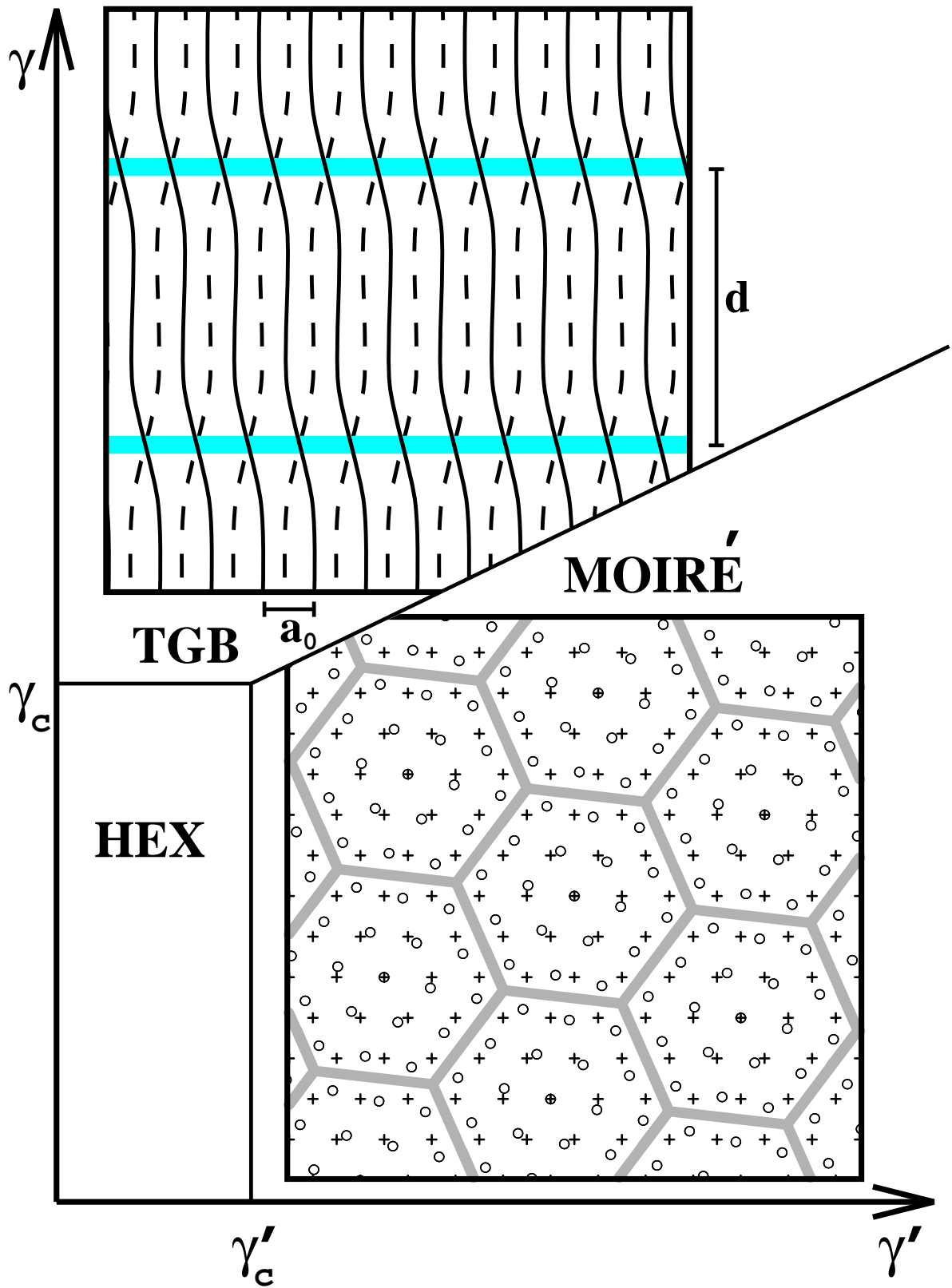
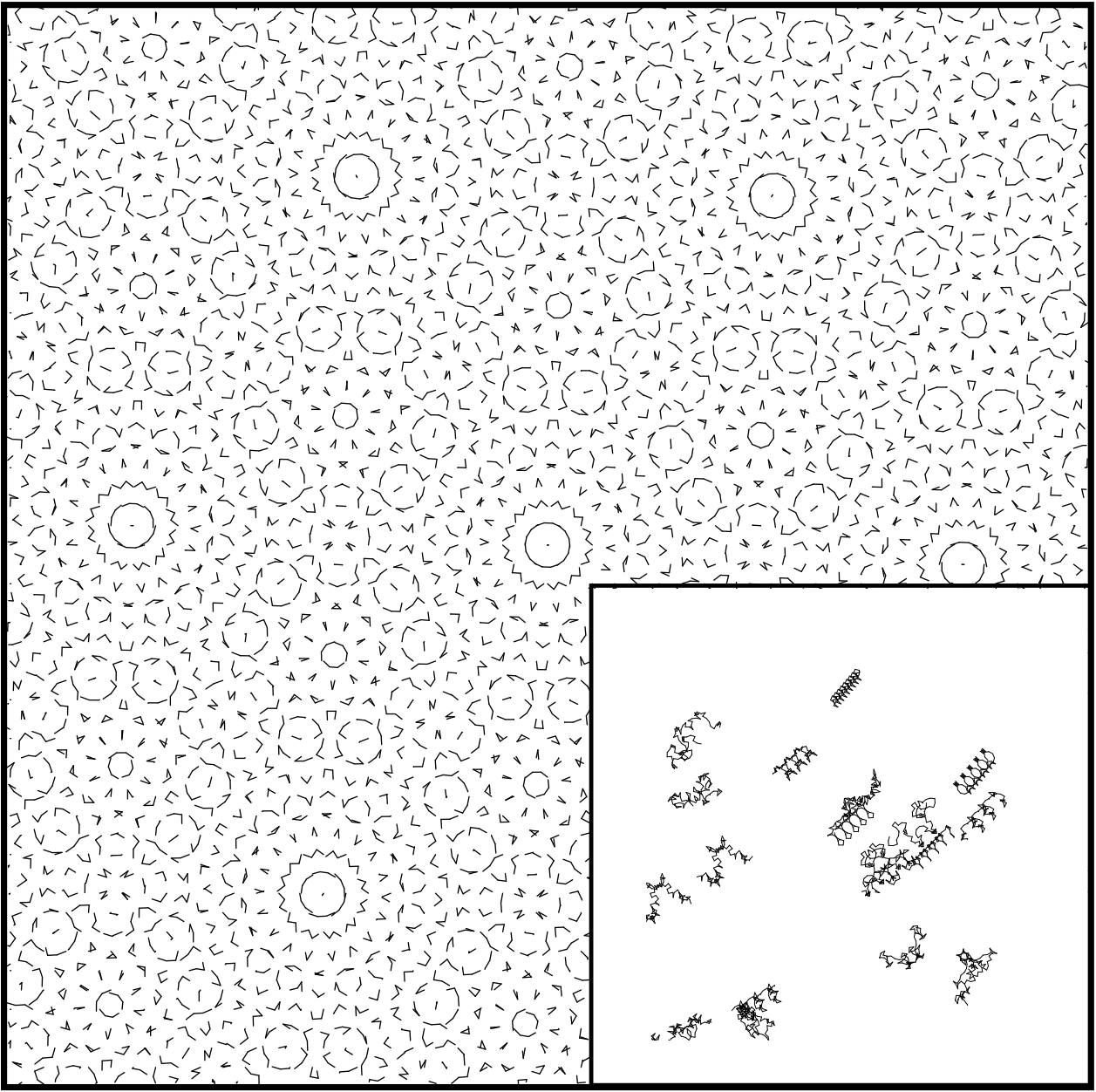
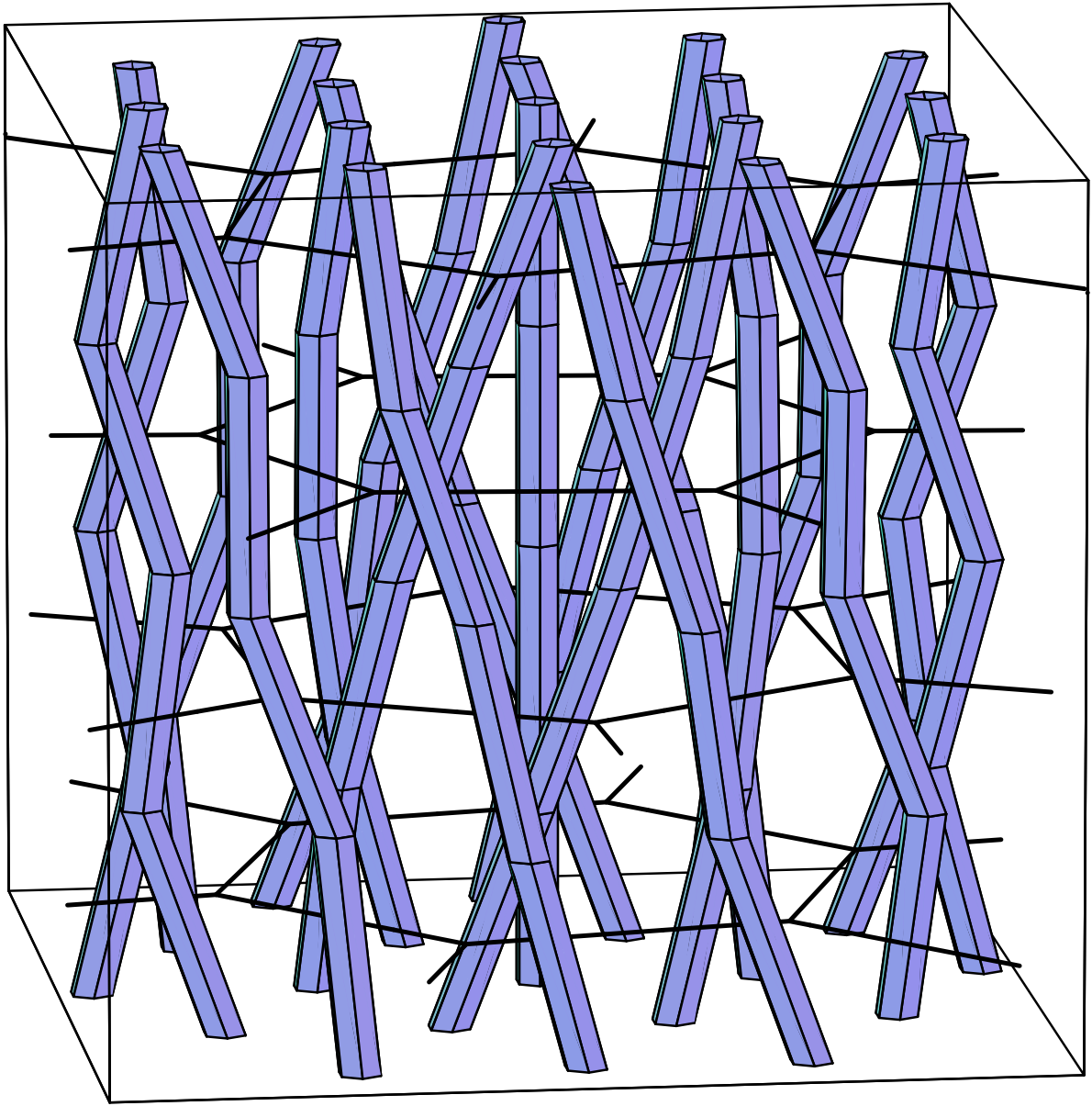
\bye